\documentclass[aps, prl, reprint, showkeys, nolongbibliography]{revtex4-1}

\pdfoutput=1

\usepackage[utf8]{inputenc}
\usepackage[usenames,dvipsnames]{color}
\usepackage{hyperref}
\usepackage{amssymb}
\usepackage{multirow}
\usepackage{epsfig}
\usepackage{rawfonts}
\usepackage{latexsym}
\usepackage{theorem}
\usepackage{natbib}
\usepackage{graphicx}

\begin{document}

\title{Nagaoka ferromagnetism in semiconductor artificial graphene}
\date{\today}

\author{G\"{o}khan \"{O}ztarhan$^{*\ddagger}$, Pawe{\l} Potasz$^\dagger$ and A. D. G\"{u}\c{c}l\"{u}$^*$}
%\email[Corresponding author.\\]{gokhanoztarhan@iyte.edu.tr}
%\email[Corresponding author.\\]{devrimguclu@iyte.edu.tr}
%\email{}
\affiliation{$^*$Department of Physics, \.{I}zmir Institute of Technology, 35430 Urla, \.{I}zmir, T\"{u}rkiye\\ 
$^\dagger$Institute of Physics, Faculty of Physics, Astronomy and Informatics, Nicolaus Copernicus University, Grudziadzka 5, 87-100 Toru\'{n}, Poland}

\begin{abstract}
We present the emergence of Nagaoka ferromagnetism in semiconductor-based artificial graphene with realistic Coulomb interaction
 using high-precision variational and diffusion Monte Carlo methods,
complemented by exact diagonalization calculations of the generalized Hubbard model. We analyze models of armchair hexagonal geometries nanopatterned on GaAs quantum wells. Our results reveal a distinct magnetic phase transition driven by the absence/addition of a single electron at half-filling. This form of itinerant magnetism predicted rigorously for Hubbard model remained unascertained in large scale realistic systems. We demonstrate that Coulomb scattering terms play a crucial
role in stabilizing Nagaoka ferromagnetism, enabling the observation
of the phase transition for system parameters near $U/t \approx 60$.
\end{abstract}


\maketitle

Nagaoka ferromagnetism, originally predicted within the
framework of the strongly correlated Hubbard model
\cite{PhysRev.147.392,Thouless_1965}, arises in the infinite U limit when a single hole in a half-filled lattice induces a fully spin-polarized state.
Intuitively, the hole motion is fully coherent in the background of a fully polarized spin configuration, minimizing the kinetic energy, in contrast to unpolarized spins that lead to many incoherent paths.
Some attempts to generalize Nagaoka’s result lead to a generalized version of Nagaoka’s theorem by Tasaki \cite{PhysRevB.40.9192} or the flat-band ferromagnetism of Mielke \cite{Mielke_1991}. Although mathematically rigorous, the Nagaoka theorem and its extensions have limited practical utility in conventional materials, where disorder and competing interactions complicate the picture. To date, suitable conditions for the observation of Nagaoka ferromagnetism have been achieved in a relatively small system—a $2\times2$ quantum dot plaquette formed electrostatically \cite{Dehollain2020}, while Nagaoka polarons were observed in a triangular optical lattice \cite{Lebrat2024}. 
In previous works, Nagaoka ferromagnetism was investigated within Hubbard model for a single hole for finite \cite{PhysRevLett.64.475,PhysRevB.64.024411} and infinite-U limit \cite{PhysRevLett.86.3396,PhysRevLett.108.126406,PhysRevB.41.2375}, and with different hole 
densities \cite{PhysRevB.77.035107, PhysRevB.40.7406}, extended Hubbard model \cite{PhysRevB.106.L041104, PhysRevB.109.235128, PhysRevB.53.9225} 
and ab initio exact diagonalization 
\cite{PhysRevB.100.155133} for small size systems.
Whether Nagaoka ferromagnetism can be realized in larger scale solid-state systems remains an open question as Coulomb interactions include not only finite on-site repulsion but also long-range and scattering terms \cite{morales2022nonlocal}. 

Concurrently, solid-state quantum simulators \cite{Buluta2009,Salfi2016,Cai2013} have become an important tool for investigating strongly
correlated electron systems \cite{li21,mak2022semiconductor}, offering unprecedented control over lattice
geometries \cite{Kennes2021}, interactions \cite{Tang2020}, and external fields, capabilities that remain
challenging to achieve in real materials
\cite{AG_low_disorder,AG_observation_of_dirac_bands,Morales-Duran2024}. Among solid-state quantum simulators, artificial graphene (AG), recently realized in quantum dot arrays using modulation-doped AlGaAs/GaAs quantum wells
\cite{Jackmin2014,Wang2016_AG,N_dvorn_k_2012},
has emerged as a powerful platform for exploring tunable Dirac physics \cite{Tarruell2012,AG_theoretical_0,AG_theoretical_1,AG_theoretical_3}
and interaction-driven phenomena \cite{Singha2011,AG_theoretical_2,AG_theoretical_yasser}. The Hubbard parameter in semiconductor AG systems is predicted to be of the order of $U/t\sim 100$ \cite{PhysRevB.108.L161114}, enabling the investigation of correlated insulating phases and emergent magnetism. Thus, with its highly tunable electron filling, interaction strength, and band structure, 
semiconductor AG provides an ideal platform for probing Nagaoka ferromagnetism under well-controlled conditions, potentially enabling the first direct realization of itinerant ferromagnetism in a programmable quantum system.

In this work, we predict the emergence of Nagaoka
ferromagnetism in semiconductor AG, using advanced computational techniques going beyond Hubbard model. We examine a 42-site hexagonal lattice — significantly larger than those studied in prior literature on Nagaoka ferromagnetism, and electrons interacting via long range Coulomb potential, using continuum variational Monte Carlo (VMC) \cite{qmc_cyrus_2} and diffusion
Monte Carlo (DMC) \cite{qmc_cyrus_1} methods to provide a
non-perturbative and highly accurate treatment of many-body correlations
\cite{qmc_review,qmc_devrim_0}. 
We focus on a hexagonal armchair geometry
(the zigzag edge termination leads to the appearance of edge states and additional magnetic phenomena), which
serves as an intermediary between finite-size samples and bulk graphene, where edge effects are mitigated through a confining potential \cite{PhysRevB.108.L161114}.
The chosen lattice size, 42, is smallest lattice size that can be created in hexagonal geometry with armchair edges (see supplemental materials (SM) for QMC $114$ sites results with similar behavior \cite{Supplemental}).
The system is modeled with a nearest-neighbor distance of $a=50$ nm, in alignment with
recent experimental findings \cite{AG_observation_of_dirac_bands} and computational studies
\cite{PhysRevB.108.L161114}.
Our results reveal a transition from an antiferromagnetic (AFM) to a
ferromagnetic (FM) phase which occurs
exactly one hole/electron away from the half-filling as described in Nagaoka
theorem \cite{PhysRev.147.392}. Our predictions are supported by appropriate exact diagonalization (ED) calculations of
the Hubbard model on a hexagonal armchair lattice. Given
the prohibitive computational complexity associated with the full Hilbert space for one hole added to the half-filling,
the ED calculations are restricted to the subspace corresponding to a single spin flip from a fully spin polarized state. The ED results further suggest that  Coulomb interaction scattering terms are necessary to
observe Nagaoka ferromagnetism for realistic system parameters.
In particular, the ferromagnetic phase is further stabilized with the exchange terms in agreement with Ref.\ \cite{PhysRevB.53.9225}.

Our model of nanostructured semiconductor AG consists of $N_{e}$
interacting electrons in a honeycomb array of $N$ confining
potentials, described by the many-body Hamiltonian in effective atomic units
(electronic charge $e$, dielectric constant $\epsilon$, effective mass $m^*$,
and $\hbar$ are set to 1),
\begin{equation} \label{hamiltonian}
  H = -\frac{1}{2} \sum_{i}^{N_{e}} \nabla^{2}_{i} + \sum_{i}^{N_{e}} V(\textbf{r}_{i})
      + \sum_{i}^{N_{e}} k |\textbf{r}_{i}|^2 + \sum_{i<j}^{N_{e}} \frac{1}{r_{ij}}
\end{equation}
where $1 / r_{ij}$ is the Coulomb interaction between the electrons,
$V(\textbf{r}_{i})$ is the total potential energy of the confining
potentials, and $k$ is the spring constant of quadratic gate potential
located at the center of the system which controls the finite size
effects. Typical material properties for GaAs, effective electron mass
$m^{*} = 0.067 m_{0}$ and dielectric constant $\epsilon = 12.4$, are
used.  Corresponding effective Bohr radius is $a^{*}_{0} = 9.794$ nm,
and the effective Hartree energy is $11.857$ meV. The honeycomb array
of potential wells is modeled using gaussian-like
functions \cite{AG_theoretical_3,PhysRevB.108.L161114},
\begin{equation} \label{potential}
    V(\textbf{r}) = V_{0} \sum\limits_{\textbf{R}_{0}}
    \exp[-(|\textbf{r}-\textbf{R}_{0}|^{2} / \rho^{2})^{s}]
\end{equation}
where $V_{0}$ is the potential depth, $\rho$ is the radius
and $s$ is the sharpness of the potential wells. $\textbf{R}_{0}$ is
the location of the potential wells. In our numerical calculations,
dot-to-dot distance was fixed to $a=50$ nm. The sharpness value
$s$ was fixed, as well, to $s = 1.4$ which is the intermediate point between
gaussian ($s = 1$) and muffin-tin like ($s = 2.8$) potentials (The results are expected to remain robust under $s$ parameter variation \cite{PhysRevB.108.L161114}). The spring constant of the quadratic gate potential
was kept at $k = 3.56 \times 10^{-4}$ meV/nm$^2$ to reduce finite size effects \cite{PhysRevB.108.L161114}.
In contrast to our previous work, where charge neutrality was maintained and the total energy was constrained near zero by tuning the confining potential $V_{0}$ \cite{PhysRevB.108.L161114}, the present study relaxes this restriction and instead employs a deeper confining potential $V_{0}$
to induce Nagaoka ferromagnetism.

Our VMC and DMC calculations were performed using Slater-Jastrow trial wave functions constructed from three different types of orbitals: (i) Gaussian functions on the sites, describing localized states, (ii) Tight-binding (TB) orbitals, suitable for metallic phases with delocalized electrons, and (iii) Mean-field Hubbard (MFH) orbitals, capable of describing both localized and liquid-like states depending on $U_{t}/t$ (see a section "Details of QMC calculations" in SM \cite{Supplemental}). 
%Note that TB and MFH orbitals consist of a linear combination of Gaussian functions in which the coefficients are obtained from the diagonalization of TB and MFH Hamiltonians.
The orbital type yielding the best fixed-node DMC energy for a given set of system parameters was chosen for further numerical analysis. The details of our VMC and DMC approach can be found in Ref.\ \onlinecite{PhysRevB.108.L161114}.

\begin{figure}[h]
\centering
\includegraphics[width=0.850\columnwidth]{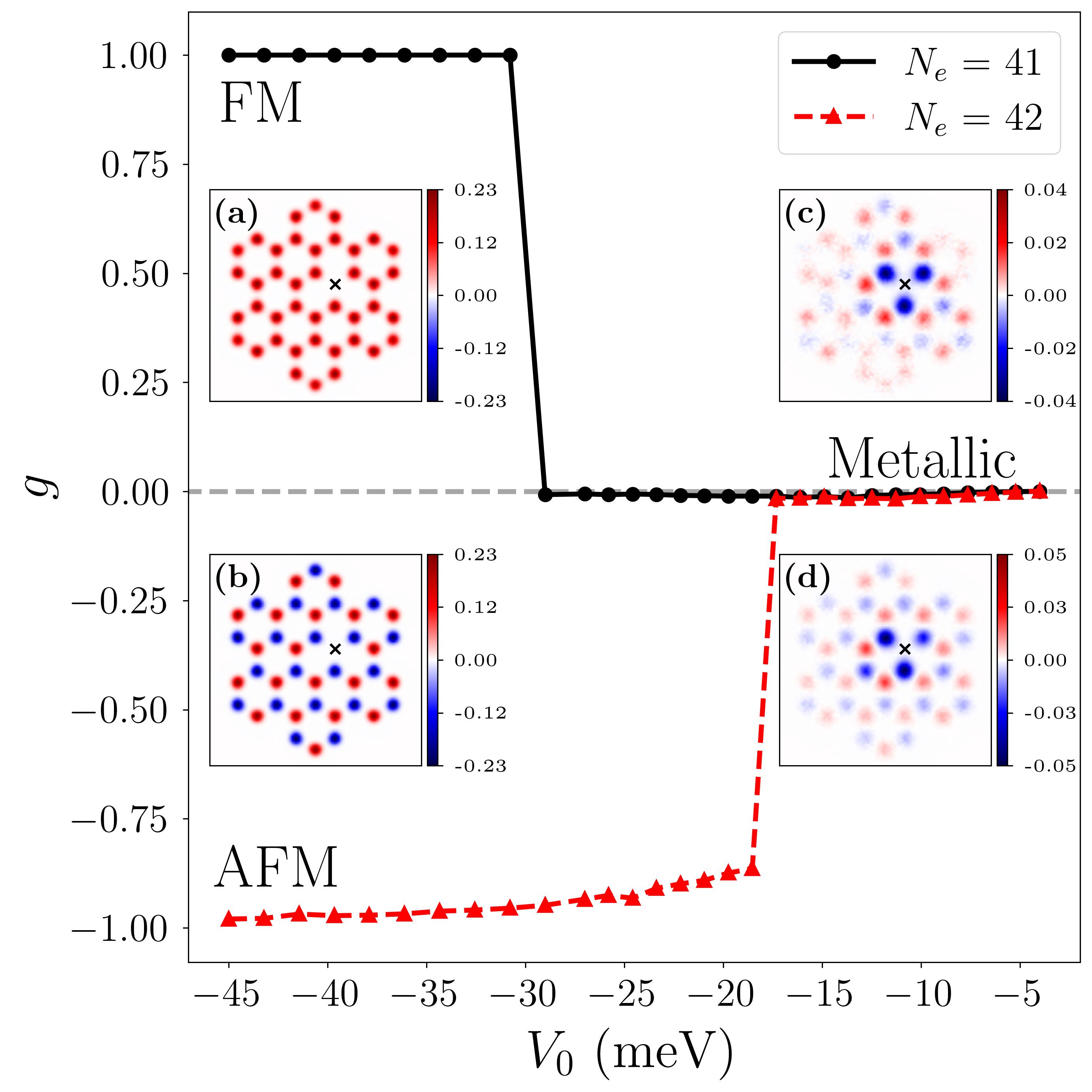}
\caption{\label{figure1}
    Extrapolated spin-spin correlation function plotted against
    potential depth $V_{0}$ obtained using pair densities for
    potential radius $\rho = 25$ nm.
    Inset figures are extrapolated spin pair densities in the ground states.
    (a) $N_{e} = 41$, $V_{0} \approx -45$ meV, $S_{z} = 41/2$.
    (b) $N_{e} = 42$, $V_{0} \approx -45$ meV, $S_{z} = 0$.
    (c) $N_{e} = 41$, $V_{0} \approx -8.84$ meV, $S_{z} = 1/2$.
    (d) $N_{e} = 42$, $V_{0} \approx -8.84$ meV, $S_{z} = 0$.
}
\end{figure}

For a deep understanding of transition dynamics including magnetic 
and metallic phases as a function of potential well depth (which controls $U/t$)
and electron number, we consider a spin-spin
correlation function averaged over all nearest neighbor pairs $(i,j)$,
defined as $g=\langle m_i m_j \rangle / \langle n_i n_j \rangle $,
where $m_i$ and $n_i$ are the average magnetization and electron
density on site $i$ within a radius $r=a/2$. To reveal the internal
spin structure, $m_i$ and $n_i$ are calculated using the pair densities
$p_{\sigma \sigma_0}(\textbf{r}, \textbf{r}_0)$, the probability of finding an electron
with spin $\sigma$ at location $\textbf{r}$ when an electron with spin
$\sigma_0$ is fixed at location $\textbf{r}_0$.  The output values of
the spin-spin correlation function remain in the $[-1, 1]$ range,
where $g = -1$ corresponds to an AFM phase,
$g = 0$ signifies a metallic configuration, and $g = 1$ indicates a fully FM phase.
In Fig.\ \ref{figure1}, spin-spin correlation function $g$ is shown as a function
of potential depth $V_{0}$ for $N = 42$ sites, $\rho = 25$ nm and $k \ne 0$. The location of the fixed up-spin electron is denoted by a cross positioned atop a lattice site-strategically chosen to break
system symmetry while avoiding edge effects. Both
$N_{e} = 42$ and $N_{e} = 41$ electron numbers are considered for half-filling
and a single hole away from half-filling, respectively. For these calculations,
only maximum and minimum values of $S_{z}$ are considered to determine the ground
state of the system. As shown in Fig.\ \ref{figure1}, the system at the half-filling exhibits a transition
from metallic to AFM insulating phase around $V_{0} \approx -18$ meV, and the system with
$N_{e} = 41$ electrons shows a transition from metallic to FM phase around
$V_{0} \approx -30$ meV. The results suggest that the system should make a
transition to AFM phase before subtracting a single electron. If
the same figure is analyzed vertically, the system exhibits a transition from
AFM to FM phase around $V_{0} \approx -30$ meV when an electron is pulled away
from the system as it is suggested in Nagaoka theorem. After $V_{0} \approx -30$
meV, the transition maintains its stability. Although, at $\rho = 25$ nm, AG exhibits
metallic-like behavior as predicted in \cite{PhysRevB.108.L161114}, increasing
effect of potential depth $V_{0}$ opens a way to have a localized behavior for the
systems. To visualize the phases depicted above by $g$, Figs. \ref{figure1}a-b-c-d show the pair spin densities $p_{\uparrow\uparrow }( \textbf{r},\textbf{r}_0)-p_{\downarrow
  \uparrow }( \textbf{r},\textbf{r}_0)$. In Fig.\
\ref{figure1}(a), the system have a fully FM phase for $N_{e} = 41$ electrons at
$V_{0} = -45$ meV with maximum $S_{z} = 41/2$ while the system with $N_{e} = 42$
electrons at the same $V_{0}$ value Fig.\ \ref{figure1}(b) exhibits an AFM phase
at minimum $S_{z} = 0$. Additionally, at the lower extrema value of
$V_{0} \approx -8.84$ meV, Fig.\ \ref{figure1}(c) and \ref{figure1}(d) shows metallic
phases for $N_{e} = 41$ with $S_{z} = 1/2$ and $N_{e} = 42$ with $S_{z} = 0$,
respectively.

\begin{figure}[h]
\centering
\includegraphics[width=0.975\columnwidth]{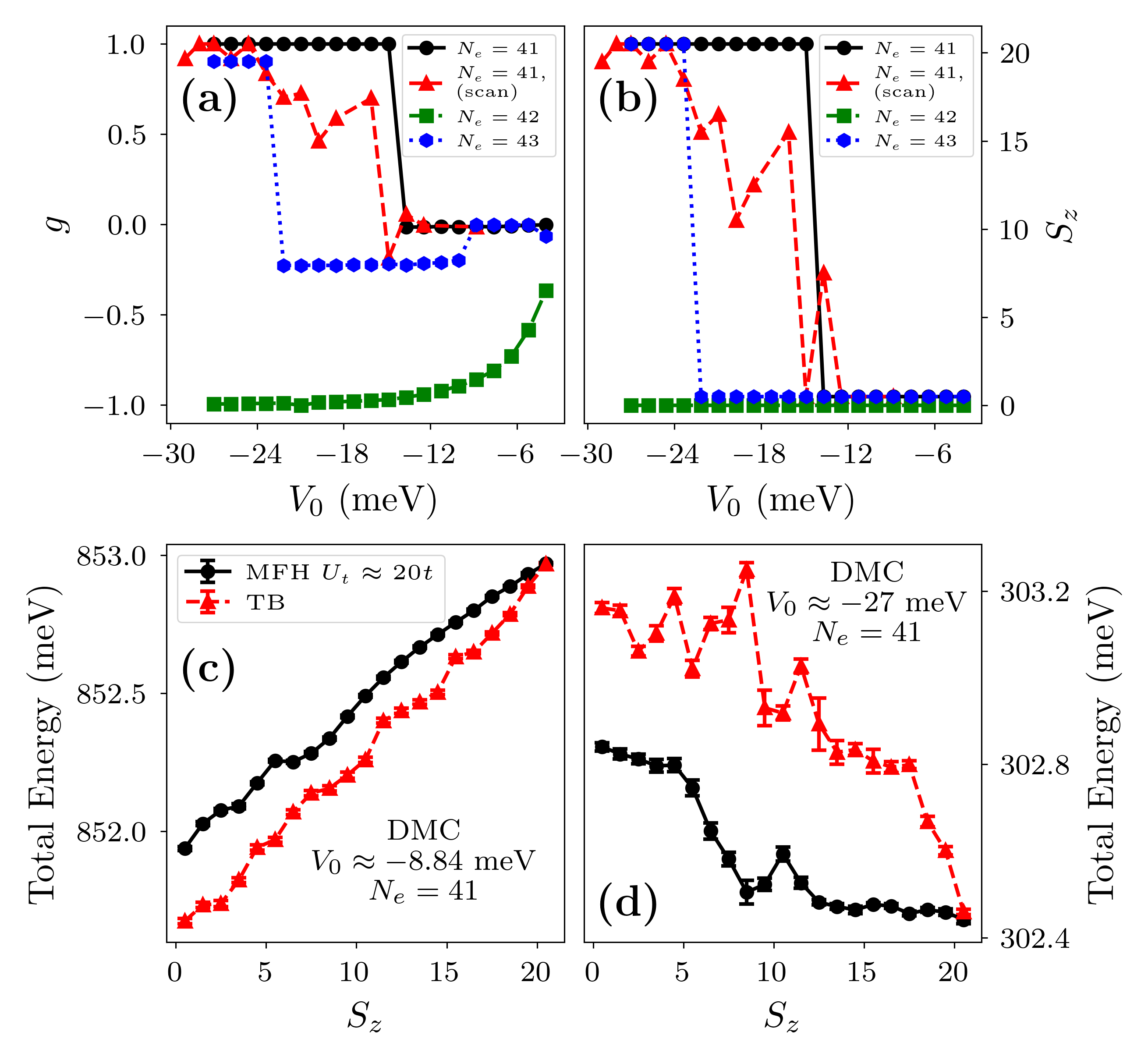}
\caption{\label{figure2}
    (a) Extrapolated spin-spin correlation function plotted against
    potential depth $V_{0}$ obtained using pair densities for
    potential radius $\rho = 17.5$ nm.
    (b) Ground state spin $S_{z}$ values plotted against potential depth $V_{0}$.
    (c) DMC energies plotted against $S_{z}$ at $V_{0} \approx -8.84$ meV for
    $N_{e} = 41$.
    (d) DMC energies plotted against $S_{z}$ at $V_{0} \approx -27$ meV for
    $N_{e} = 41$.
    For (a) and (b), the calculations are for $S_{z}$-min/max competition except "(scan)" results where we consider all $S_{z}$ values.
}
\end{figure}

In Fig.\ \ref{figure2}, we turn our attention to more localized systems
in which potential radius $\rho = 17.5$ nm, in order to strengthen the electron
correlations \cite{PhysRevB.108.L161114}. Fig.\ \ref{figure2}(a) shows spin-spin
correlation function as a function of $V_{0}$ for $N_{e} = 41$, $N_{e} = 42$ and
$N_{e} = 43$ electrons. In order to validate the results in more detail,
a full spin $S_{z}$ scan have been also performed for $N_{e} = 41$ to determine
the ground states of the system for a given $V_{0}$. In contrast to Fig.\
\ref{figure1}, half-filling system already shows an AFM behavior starting from
higher $V_{0}$ values shown in Fig. \ref{figure2}(a), which is directly correlated
with potential radius $\rho = 17.5$ nm used in these calculations. The ground
states are determined from extrema $S_{z}$ values which we denote as $S_{z}$-min/max.
From $S_{z}$-min/max calculations of $N_{e} = 41$ electrons, it is observed that
the sharp transition from metallic to FM phase is at higher value
$V_{0} \approx -15$ meV than $\rho = 25$ nm case. Thus, an AFM to FM phase
transition also occurs in $\rho = 17.5$ nm system depending on electron number
yet in a higher $V_{0}$ value effected by strong electron-electron interactions.
Full $S_{z}$-scan of
$N_{e} = 41$ electrons shows a smoother transition (apart from DMC statistical
fluctuations) starting from $V_{0} \approx -15$
meV as shown in Fig. \ref{figure2}(a)-(b). The system is in
fully polarized FM phase around $V_{0} \approx -26$ meV. 
In Fig.\ \ref{figure2}(a), we also show $N_{e} = 43$ electrons results in which
the system first exhibits a low AFM phase then exhibits a transition to FM
state around $V_{0} \approx -23$ meV. While $g$ is slightly lower than $1$ due to the additional
opposite spin electron in the system, $S_z$ shown in Fig.\ \ref{figure2}(b) indicates a fully polarized FM phase. Therefore, the AFM to FM phase transition
also occurs by adding one electron to the system, as suggested by Nagaoka theorem for an ideal honeycomb system where electron-hole symmetry is present. It is interesting to notice that before a transition to a spin polarized phase after adding one electron, a stable antiferromagnetic order is present that is absent for a one hole case. The asymmetry between $\pm$ one extra electron is expected in a frustrated triangular lattice, and here is related to violation of bipartiteness of the lattice in a realistic system. Weak metallic antiferromagnetism of kinetic origin after adding one hole to a half-filled triangular lattice was predicted by Haerter and Shastry \cite{PhysRevLett.95.087202}.

Fig.\ \ref{figure2}(c)-(d)
show DMC total energies at two examples of extrema $V_{0}$ values of
$-8.84$ meV and $-27$ meV
for $N_{e} = 41$ electrons. MFH $U_{t} \approx 20t$ trial wave functions represents
more localized states, and TB trial wave functions represents liquid-like states.
As seen from Fig.\ \ref{figure2}(c), the lowest energy is given by the TB trial
wave function at $S_{z} = 1/2$ for $V_{0} \approx -8.84$ meV.
In Fig.\ \ref{figure2}(d), the ground state is
represented by the MFH $U_{t} \approx 20t$ trial wave function at $S_{z} = 41/2$
for $V_{0} \approx -27$ meV.

\begin{figure}[h]
  \centering
  \includegraphics[width=0.850\columnwidth]{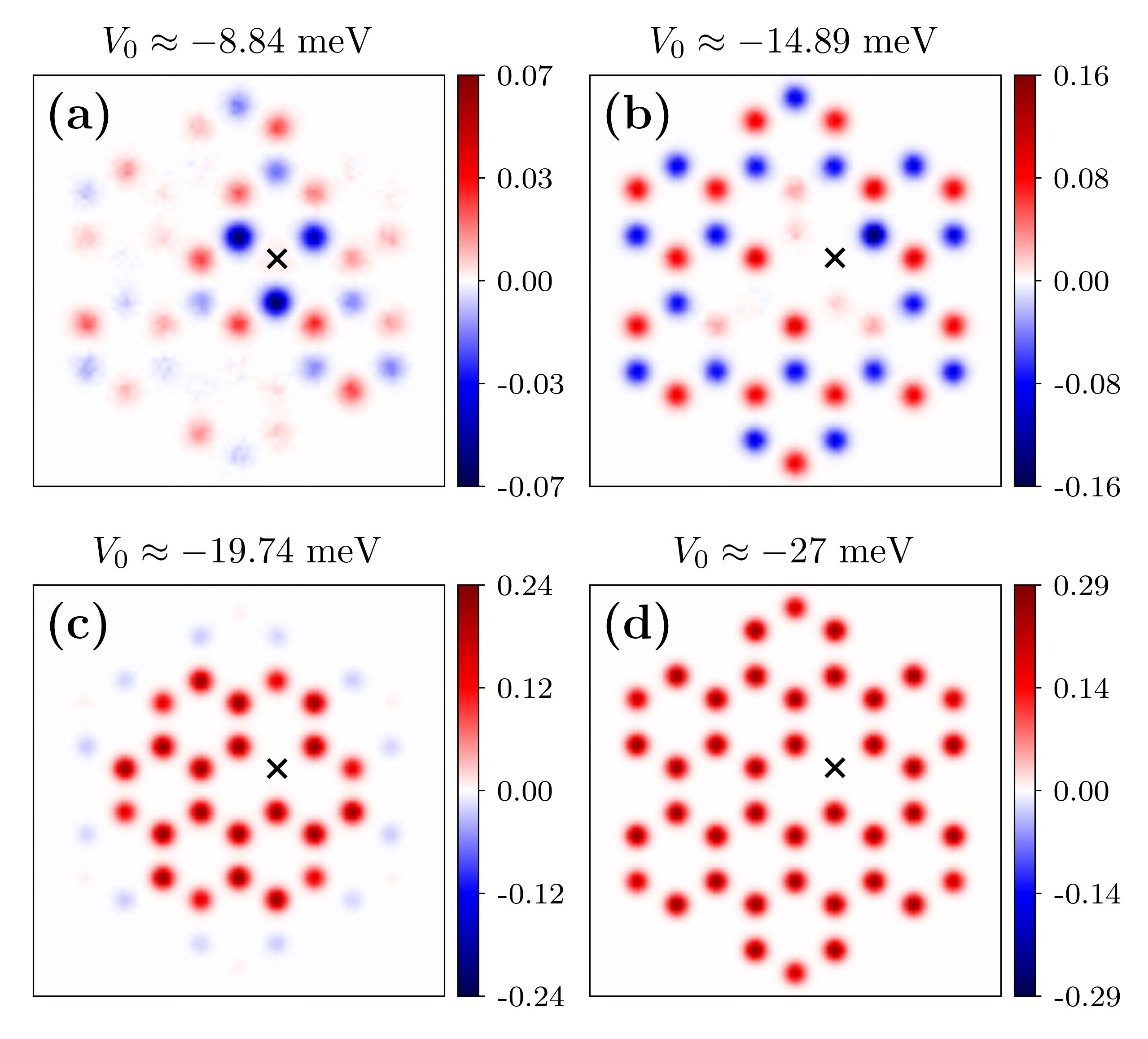}
  \caption{\label{figure3}
    Extrapolated spin pair densities for $N_{e} = 41$ electrons
    at several potential depth $V_{0}$ values.
    (a) $V_{0} \approx -8.84$ meV.
    (b) $V_{0} \approx -14.89$ meV.
    (c) $V_{0} \approx -19.74$ meV.
    (d) $V_{0} \approx -27$ meV.
  }
\end{figure}

In Fig. \ref{figure3}, extrapolated spin pair densities for $N_{e} = 41$
electrons on $N = 42$ sites are shown at several potential depth $V_{0}$ values
being chosen to reveal the nature of the phase transition. While a liquid-like
state is seen from Fig.\ \ref{figure3}a at a high $V_{0}$ value, a fully polarized
FM state is observed around $V_{0} \approx -27$ meV (Fig.\ \ref{figure3}d),
consistent with $g$ values
shown in Fig.\ \ref{figure2}a. An important point is that before going into
a phase transition to FM state, the system exhibits an AFM state around
$V_{0} \approx -14.89$ meV due to increasing electron-electron correlations, as
seen from Fig.\ \ref{figure3}b.
During the transition around $V_{0} \approx -19.74$ meV in Fig.\ \ref{figure3}c,
the polarization of spins located more around the center. This behavior hints
a possible Nagaoka polaron formation in AG systems.

\begin{figure}[h]
  \centering
  \includegraphics[width=0.80\columnwidth]{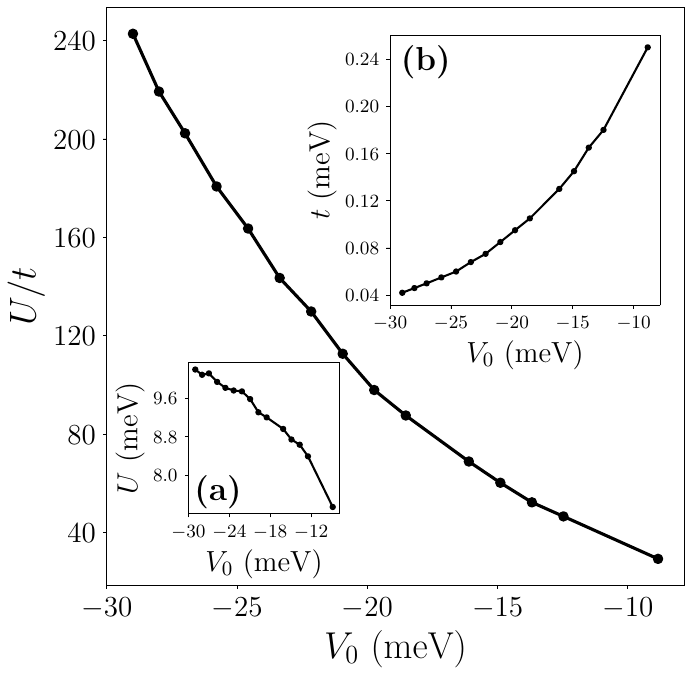}
  \caption{\label{figure4}
    The ratio of onsite Coulomb interaction $U$ to nearest-neighbor
    hopping parameter $t$, $U/t$,
    plotted against $V_{0}$ for $\rho = 17.5$ nm and $k = 3.56 \times 10^{-4}$ meV/nm$^2$.
    (a) $U$ vs. $V_{0}$.
    (b) $t$ vs. $V_{0}$.
  }
\end{figure}

In order to reveal the intrinsic strength of electron-electron interactions, the
onsite Coulomb interaction, $U$, is determined using extrapolated electron
densities from QMC calculations of $S_{z}$-scanned systems of
$N_{e} = 41$ electrons.
Conversely, the nearest-neighbor hopping
parameter, $t$, is estimated through a hybrid approach that combines solutions
of the single-particle Schr\"{o}dinger equation (SPSE) with the diagonalization of
the TB Hamiltonian for an artificial benzene molecule in which $\rho = 17.5$ nm.
The energy spectra obtained from SPSE solutions and TB Hamiltonian
diagonalization are systematically compared and fitted. The hopping parameter,
$t$, is subsequently extracted by iteratively adjusting and refining the TB model
to achieve optimal agreement with the computed spectra. As shown in Fig.\
\ref{figure4}, the ratio $U/t$ increases in an exponential fashion as
$V_{0}$ decreases. The exponential increase of $U/t$ is a direct consequence of
increasing $U$ and decreasing $t$ for decreasing $V_{0}$, shown in Fig.\
\ref{figure4}a and \ref{figure4}b. Around the starting point of AFM to FM phase
transition, $V_{0} \approx -15$, the ratio $U/t \approx 60$ while around end point,
$V_{0} \approx -26$, the ratio $U/t \approx 170$. Obtained values of the ratio $U/t$
are in good agreement with previous results within  triangular optical lattice experiments (for $U/t \approx 72$ Nagaoka polaron around a doublon covers around 30 sites) \cite{Lebrat2024} and within density matrix renormalization group calculation on a square lattice \cite{PhysRevB.64.024411}.

\begin{figure}[h]
\centering
\includegraphics[width=0.90\columnwidth]{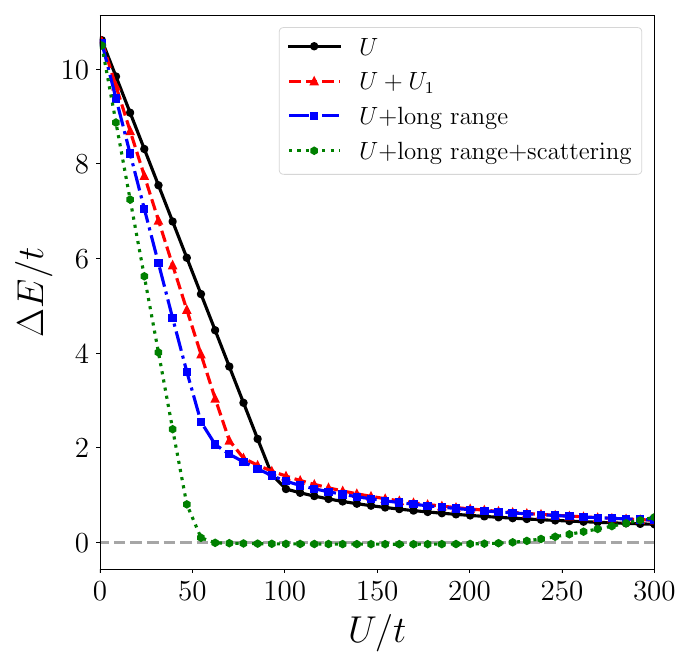}
\caption{\label{figure5}
    Exact diagonalization results of the Hubbard Hamiltonian in a subspace of
    a single spin flip on $N = 42$ sites with $N_e=41$ electrons (one hole). The energy difference $\Delta E$ between
    $S=39/2$ and $S=41/2$ states' lowest energy states. Confining potential strength $\omega = 0.5 t / a^{2}$
    where $a$ is the dot-to-dot distance.
}
\end{figure}

QMC results can be supported by ED calculations in a vicinity of fully spin polarization subspace. ED suffers from an exponential growth of many-body Hilbert space with a system size. For $N=42$ lattice size system in the vicinity of the half-filling, this corresponds to $\sim 10^{23}$ basis states for minimum $S_{z}$ subspace, not accessible within current computational resources. However, one can analyze the stability of a fully spin polarized state from maximum $S_{z}$ subspace ($S^{max}_{z}$) with respect to a spin flip, an energy difference between minimum energies with total spin states with $S=S^{max}-1$ and $S=S^{max}$, $\Delta E=E_{\rm min}(S^{max})-E_{\rm min}(S^{max}-1)$. We consider a hexagonal
armchair AG flake consisting of $N=42$ sites in the presence of a quadratic
confining potential of strength $\omega = 0.5 t / a^{2}$, where $a$ is a
dot-to-dot distance, a model that mimics QMC calculation in a continuum real space. These computations are
carried out within subspaces corresponding to $41$ and $40$ down-spin
electrons for a fixed total electron number $N_{e}=41$. ED calculations on a lattice allow us to analyze the role of different Coulomb interaction terms that are all present in  real space QMC simulation.  In Fig. \ref{figure5}, the energy differences between $S=39/2$ and $S=41/2$ states
are presented for four different Hamiltonians with respect to $U/t$ ratio, where
$U$ is the onsite Coulomb interaction, $U_{1}$ is the nearest neighbor direct
Coulomb interaction, long range consists of all neighbor direct Coulomb interactions, and
scattering terms includes assisted-hopping $A$, pair-hopping and exchange terms $X$ (see SM for details \cite{Supplemental}).
$U_{1}$ is calculated from QMC results as $0.24 U$, $A$ is taken as $0.2U$, and
the exchange term $X=0.05U$ as in real graphene \cite{GrapheneQuantumDots} (these values are also within a parameter range estimated for moire superlattices \cite{PhysRevB.108.165152}). 
For all cases, there is dramatic decrease in the energy difference around $U/t \approx 70$,
which is consistent with QMC results. The behavior of energy differences as a function of $U/t$ ratio for only $U$, $U + U_{1}$ and $U +$ long range cases are quite similar, but the transition to a fully spin polarized state is not expected within a range of $U/t$ ratio shown in a figure. However, when scattering terms are added, the energy difference is lowered by the order of $t$, and becomes negative.
The maximum polarized $S=41/2$ state has the lowest energy around $U/t \approx 60$. This feature is more general regardless of a choice of given parameters, see SM \cite{Supplemental}. Stabilization of the maximum polarized state by adding scattering terms (especially the exchange terms) is consistent with the findings in Ref.\ \cite{PhysRevB.53.9225}.
This hints a possible transition to FM phase around similar $U/t$ ratio as in
the QMC results which verifies the phase transition predicted by QMC.

In summary, we demonstrated that Nagaoka ferromagnetism can be probed in
semiconductor artificial graphene using accurate variational
and diffusion Monte Carlo calculations, alongside exact diagonalization of
the Hubbard model. Specifically, quantum Monte Carlo calculations 
on $N = 42$ sites near half-filling reveal an antiferromagnetic to
ferromagnetic phase transition driven by the absence of a single electron.
Additionally, the phase transition can be induced by adding an electron to the half-filling,
consistent with Nagaoka's theorem on bipartite lattices. As $V_{0}$ decreases
(i.e. $U/t$ increases), the $N_{e} = 41$ electron
system evolves from a liquid-like state to a fully ferromagnetic phase, passing
through an intermediate antiferromagnetic phase where spin polarization gradually
builds near the center of system, suggesting possible Nagaoka polaron formation.
We estimate the $U/t$ ratio and determine that the phase transition occurs above
$U/t \approx 60$, indicating that strong electron correlation is necessary. 
Furthermore, exact
diagonalization of the generalized Hubbard model, constrained to near-maximal polarization
due to computational limitations, confirms the phase transition.
Adding assisted-hopping and exchange interaction terms to the Hamiltonian
stabilizes the transition that validates the $U/t$ ratio predictions of
quantum Monte Carlo calculations. 
We note that while QMC calculations were performed for GaAs quantum dots parametrization, similar lattices can also be created in moir\'e superlattice \cite{Yankowitz2012}. Recent studies suggested observation of graphene like energy bands, Dirac fermions, in transition metal dichalcogenides (TMD) homobilayers, in particular in so called $\Gamma$ valley TMDs \cite{PNAS_Angeli,Vitale_2021,PhysRevB.103.155142}. The bandwidth can be controlled by a twist angle and can vary from 250 to a few meV. An advantage of moire superlattices over artificial graphene superlattices made of GaAs is an ease in parameter tuning of the models. Our artificial graphene GaAs superlattice parameters would correspond to tTMD homobilayers with a twist angle between 1 and 2. 

Thus, our findings establish the presence of an 
antiferromagnetic to ferromagnetic phase transition triggered by a single electron removal/addition in a solid-state system within a realistic Hamiltonian model fully 
accounting for long-range and scattering interactions. These results suggest potential applications in nanoscale spintronic devices utilizing controllable ferromagnetism, as well as in quantum simulation platforms.

\begin{acknowledgments}
  The quantum Monte Carlo calculations reported in this study were performed using the CHAMP program \cite{Cha-PROG-XX} and were partially carried out at TUBITAK ULAKBIM, High Performance and Grid Computing Center (TRUBA resources).\\
\end{acknowledgments}

% Add corresponding author email
{\small
$^\ddagger$Email: \href{mailto:gokhanoztarhan@iyte.edu.tr}{gokhanoztarhan@iyte.edu.tr}
}

\bibliography{main}

\end{document}

\title{Supplemental Materials for ``Nagaoka ferromagnetism in semiconductor artificial graphene''}
\date{\today}

\author{G\"{o}khan \"{O}ztarhan$^*$, Pawe{\l} Potasz$^\dagger$ and A. D. G\"{u}\c{c}l\"{u}$^*$}
%\email[Corresponding author.\\]{gokhanoztarhan@iyte.edu.tr}
%\email[Corresponding author.\\]{devrimguclu@iyte.edu.tr}
%\email{}
\affiliation{$^*$Department of Physics, \.{I}zmir Institute of Technology, 35430 Urla, \.{I}zmir, Turkey}
\affiliation{$^\dagger$Institute of Physics, Faculty of Physics, Astronomy and Informatics, Nicolaus Copernicus University, Grudziadzka 5, 87-100 Toru\'{n}, Poland}

\maketitle

\section{Exact diagonalization Method}
The model we study has a single orbital per a lattice site that can be populated by up to two particles with opposite spins. The Hilbert space can be divided into smaller subspaces with total spin $S$ and azimuthal spin $S_{\rm z}$. The basis is constructed in an occupation number representation, distributing particles among single-particle states
labeled by $S_{\rm z}$. 
The total number of possible configurations $N_{\rm st}$ for particles distributed on $N$ single particle states with a given spin $N_\downarrow$ or $N_\uparrow$ is determined by a product of binomial coefficients,  $N_{\rm st}=\binom{N}{N_{\downarrow}} \cdot \binom{N}{N_{\uparrow}} $. 
The many-body Hamiltonian is diagonalized in 
$S_{\rm z}$ subspaces. We do not rotate the Hamiltonian matrix to a $S$ basis as this is an additional computational cost, and instead determine the ground state properties from calculations of expectation value of total spin $S$ for each energy eigenstate. For the analysis of the Nagaoka ferromagnetism stability, we perform calculations with $N=42$ lattice sites, $N_\downarrow = 40$ or $N_\uparrow = 1$, corresponding to $N_{\rm st}= 35301$.

Four-center real-space Coulomb matrix elements are 
\begin{equation}
\label{Coulomb}
    \langle ij|V|kl\rangle=\frac{e^2}{4\pi\epsilon_0 \epsilon} \int\int d{\bf r} d{\bf r'} \frac{\phi^*({\bf r}-{\bf R}_i) \phi^*({\bf r}'-{\bf R}_j) \phi({\bf r}'-{\bf R}_k) \phi({\bf r}-{\bf R}_l)}{|{\bf r}-{\bf r}'|},
\end{equation}
with $e$ as electric charge and $\epsilon_0$ is the vacuum permittivity, $\epsilon$ is static dielectric screening,  ${\bf r}({\bf r}')$ coordinates of  particles, and ${\bf R}_i$ is a position of $i$-th lattice site, and two-center integrals are: onsite Hubbard interaction $U_0 = \langle ii|V|ii\rangle$, nearest neighbor direct  interaction $U_1 = \langle ij|V|ji\rangle$, exchange  interaction $X_1 =\langle ij|V|ij\rangle$, pair hopping $P_1 =\langle ii|V|jj\rangle$, assisted hopping $A = \langle ii|V|ij\rangle$, where in all these terms $i,j$ are nearest neighbors and $U_n = \langle ij|V|ji\rangle$ for $j$ labeling $n$-th nearest neighbor to site $i$. We take $U_n = U_1/r_n$, where $r_n$ is a distance to $n$-th nearest neighbors, assuming $r_1=1$ for the nearest neighbors. For real wave functions $\phi = \phi^*$, $P_1 = X_1$.

\subsection{The effect of a confining potential}
\begin{figure}[h]
\centering
\includegraphics[width=0.550\columnwidth]{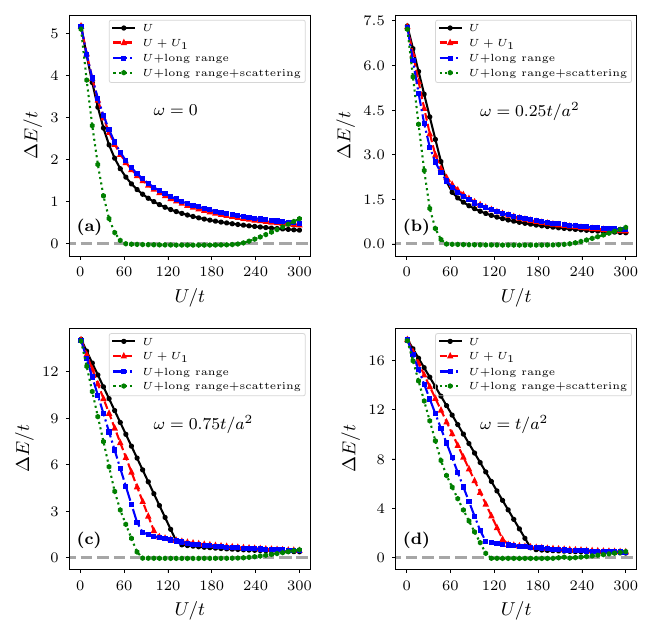}
\caption{\label{figuresm1}
    Exact diagonalization results of the Hubbard Hamiltonian in a sub-space of
    a single spin flip on $N = 42$ sites: the energy difference between
    $S=39/2$ and $S=41/2$ states. Confining potential strength $\omega$ comparison.
    (a) $\omega = 0$. (b) $\omega = 0.25 t / a^{2}$.
    (c) $\omega = 0.75 t / a^{2}$. (d) $\omega = t / a^{2}$.
    $a$ is the dot-to-dot distance.
}
\end{figure}

In Fig. \ref{figuresm1}, we show the results for different values of a confining potential, in Fig. \ref{figuresm1}(a) $\omega = 0$. \ref{figuresm1}(b) $\omega = 0.25 t / a^{2}$, \ref{figuresm1}(c) $\omega = 0.75 t / a^{2}$, \ref{figuresm1}(d) $\omega = t / a^{2}$ with $U_{1} = 0.24 U$, $A =0.2U$, and $X=0.05U$, the interaction parameters used in the main article. In all considered cases, an indication of a transition to Nagaoka ferromagnetism is observed for a model with scattering interaction terms. For a sufficiently strong confining potential, $\omega \geq 0.75 t / a^{2} $, long range interaction decreases $\Delta E$ in comparison to the models with short range interaction.

\begin{figure}[h]
\centering
\includegraphics[width=0.350\columnwidth]{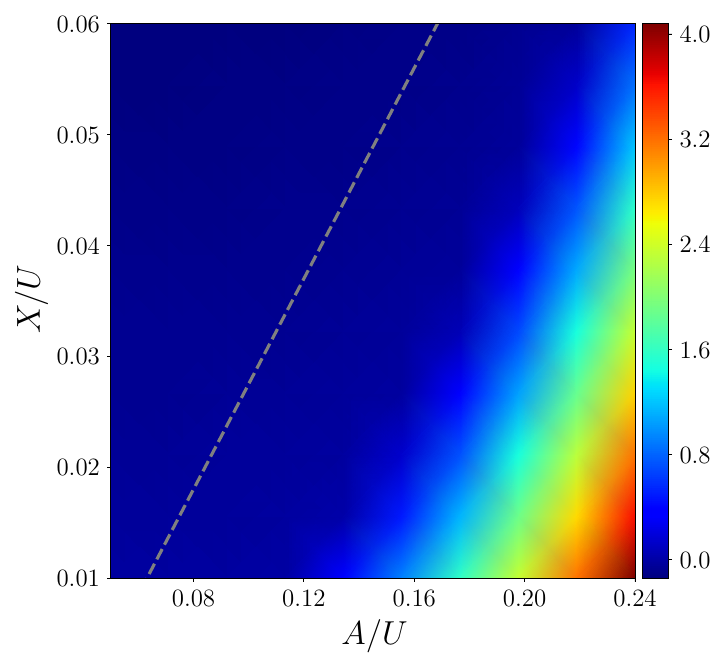}
\caption{\label{figuresm2}
    Exact diagonalization results of the Hubbard Hamiltonian in a subspace of
    a single spin flip on $N = 42$ sites: the energy difference $\Delta E$ between
    $S=39/2$ and $S=41/2$ states. Confining potential strength $\omega = 0.5 t / a^{2}$
    where $a$ is the dot-to-dot distance. $A$ is assisted hopping, and $X$ is the
    exchange term in terms of onsite Coulomb interaction $U$. The results for $U/t = 200$. Above  the gray dashed line the energy difference is negative meaning  that the lowest energy corresponds to the maximum spin polarized state.
}
\end{figure}

While $U$ and $U_1$ can be accurately calculated and are determined through the shape of a confining potential and a dot-to-dot distance, respectively, scattering interaction terms are sensitive to a particular form of wave functions. In Fig. \ref{figuresm2}, a map of $\Delta E$ is shown with $A$ on a horizontal and $X$ on a vertical axis for $U/t = 200$. A dashed line approximately indicates a transition to a region with a fully spin polarized ground state (dark blue area). This confirms that the results shown in the main article are not related to a particular choice of the parameters. While $X$ supports a transition to Nagaoka ferromagnetism, $A$ works against it. 

\begin{figure}[h]
\centering
\includegraphics[width=0.650\columnwidth]{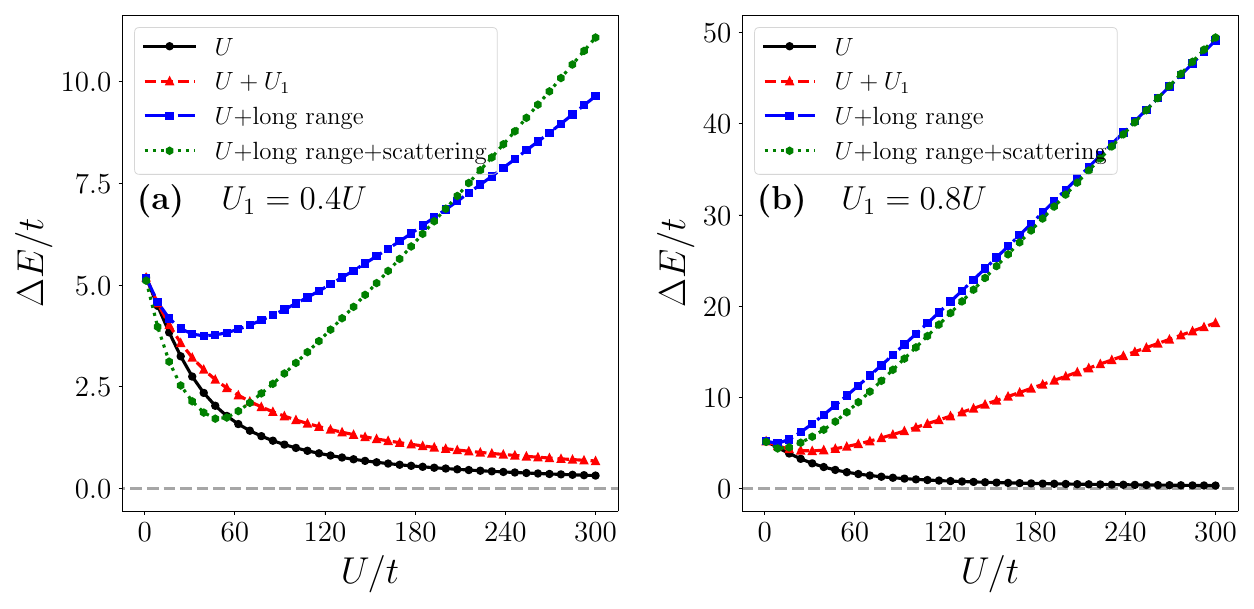}
\caption{\label{figuresm3}
    Exact diagonalization results of the Hubbard Hamiltonian in a subspace of
    a single spin flip on $N = 42$ sites: the energy difference between
    $S=39/2$ and $S=41/2$ states. Confining potential strength $\omega = 0$.
    (a) $U_{1} = 0.4 U$. (b) $U_{1} = 0.8 U$.
}
\end{figure}

In Fig. \ref{figuresm3}, we show the results of $\Delta E$ for $U_{1} = 0.4 U$ in (a) and for $U_{1} = 0.8 U$ in (b) without a confining potential, $\omega = 0$. Here a transition to Nagaoka ferromagnetism is not expected. The long range interaction works against the transition to the fully spin polarized state.  

\newpage

\section{Details of QMC calculations}
In QMC calculations, electrons may escape from the confining potential. For example, starting with four electrons in the system, one might end up with only three, as the fourth tends to drift toward infinity due to electron–electron interactions. To prevent this, the site potential, controlled by $V_0$, must be chosen sufficiently deep. This condition is reflected in the total energy of the system: if the total energy is negative, the electrons remain bound. In the present work, this issue does not arise, as $V_0$
is made deeper to induce Nagaoka ferromagnetism. In realistic artificial graphene systems, an overall gate potential can be applied, providing experimental control over the electron number in the system.

We consider finite-size systems with 42 and 114 sites, without periodic boundary conditions. In fact, VMC/DMC simulations are performed with open boundaries. To reduce edge effects, we include a quadratic confining potential, controlled by a parameter $k$, in the Hamiltonian given by Eq.\ (1). Without such a potential, electrons tend to maximize their separation due to the long-range Coulomb repulsion, which prevents meaningful extrapolation to larger system sizes, including the infinite periodic case. The confining potential helps to mimic an extended system, approaching the thermodynamic limit. In graphene lattices, two stable edge terminations are common: zigzag and armchair. Zigzag edges host localized edge states, which we aim to avoid, since our goal is to infer bulk properties in the thermodynamic limit from finite-size calculations. Therefore, we adopt the armchair edge termination.

\subsection{Trial wavefunctions}
In our study, we employ three types of trial orbitals: localized Gaussian functions, and linear combinations of localized Gaussian functions with coefficients obtained from diagonalization of tight-binding (TB) and mean-field Hubbard Hamiltonians with $U_{t}=20t$. The choice of $U_{t} = 20t$ provides an intermediate trial wave function between TB and Gaussian orbitals. Comparisons among these orbital types are presented in Ref. 45 and in Fig. 2 of the present manuscript. Since no single orbital type is universally optimal across all system parameters, we select, for each case, the orbitals that yield the lowest energy. The optimal choice depends on the system: for metallic states, TB orbitals generally perform better, whereas for localized states, mean-field Hubbard orbitals are typically more accurate. Finally, we tested Gaussian-like orbitals with oscillating tails as an alternative to Gaussian orbitals, but found no improvement in the results.

\section{QMC extra results}

\subsection{Variational Monte Carlo optimization process and DMC results}
\begin{figure}[h]
\centering
\includegraphics[width=0.750\columnwidth]{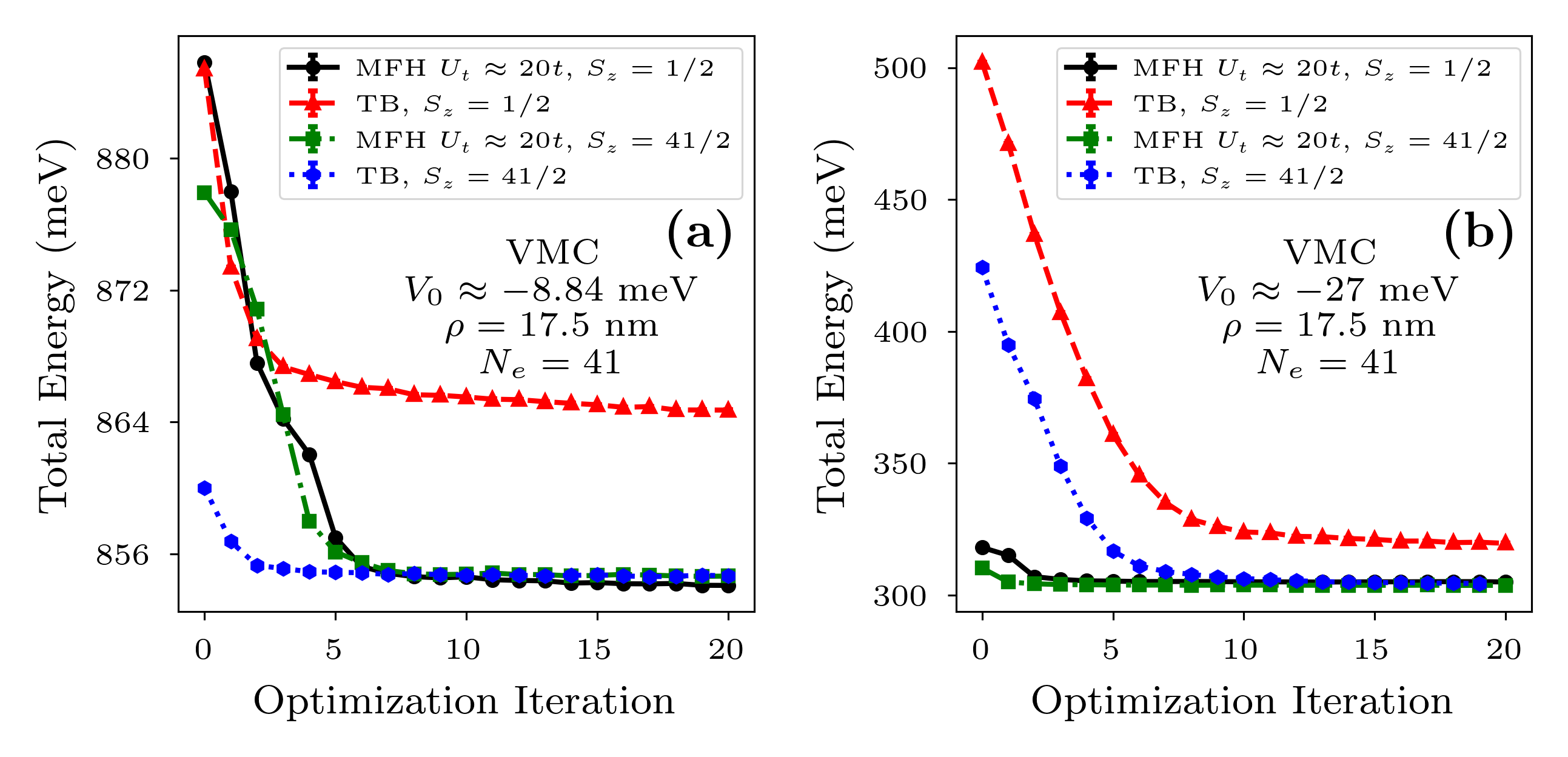}
\caption{\label{figuresm5}
    VMC total energy vs. optimization iteration plotted for two extrema $V_{0}$ values for $\rho=17.5$ nm and $N_{e}=41$ electrons.
    (a) $V_{0}\approx-8.84$ meV.
    (b) $V_{0}\approx-27$ meV.
}
\end{figure}

We have constructed Slater-Jastrow trial wave functions for the optimization part of  
VMC calculations. Slater determinant part of the wave functions consists of three 
orbital types. Gaussian functions defined on lattice sites, linear combination of 
gaussian functions by diagonalization of TB and MFH $U_{t} \approx 20t$ Hamiltonians. 
For all $V_{0}$ values, calculations are obtained from all types of orbitals. Widths 
and the positions of the gaussian functions (i.e. the nodal structure of wave 
functions), and the Jastrow parameters are optimized through VMC calculations in 
order to obtain a reasonable starting point for DMC. In Fig.\ \ref{figuresm5}, total 
energies are plotted against VMC optimization iteration for two types of wave 
functions. Two extrema values are chosen from $V_{0}$ values for $\rho=17.5$ nm and 
$N_{e}=41$ electrons. Fig.\ \ref{figuresm5}(a) shows $V_{0}\approx -8.84$ meV 
calculations in which the convergence is obtained after $7$ optimization steps for 
all types of orbitals and $S_{z}$ values. The lowest VMC energy is repsented by MFH 
$U_{t} \approx 20t$, $S_{z}=1/2$ for $V_{0}\approx -8.84$ meV. A similar picture is 
observed from Fig.\ \ref{figuresm5}(b). The convergence is about $8$ optimization 
steps, and the lowest VMC energy is represented by MFH $U_{t} \approx 20t$, $S_{z}=41/2$ for 
$V_{0}\approx -27$ meV.

The ground state of the system is revealed by DMC after VMC calculations. The lowest 
DMC energy is chosen to represent the ground state of the system at the corresponding 
$V_{0}$ and $S_{z}$ value. 
In DMC calculations, the time step was fixed at $\tau = 0.1$ after testing for time step errors. We employed 56 process per point (i.e. orbital type, $S_{z}$ and $V_{0}$), and
a target number of walkers of 500 per process to obtain a desired balance between statistical and population control error.
In Table \ref{tablesm1}, the DMC energies are shown. The 
ground state of the system at $V_{0}\approx -8.84$ is represented by TB, $S_{z}=1/2$.
For $V_{0}\approx -27$, the chosen wave function for the ground state is MFH $U_{t} \approx 20t$, $S_{z}=41/2$.

\ \

\begin{table}[h]
\centering
\begin{tabular}{|c|c|c|c|}
\hline
orbital type&$S_{z}$&$V_{0}$&total DMC energy \\ \hline
MFH&41/2&-27&302.44 $\pm$ 0.05 \\ \hline
TB&41/2&-27&302.52 $\pm$ 0.02 \\ \hline
MFH&1/2&-27&302.96 $\pm$ 0.03 \\ \hline
TB&1/2&-27&303.07 $\pm$ 0.12 \\ \hline
TB&1/2&-8.84&851.68 $\pm$ 0.14 \\ \hline
MFH&1/2&-8.84&851.94 $\pm$ 0.09 \\ \hline
TB&41/2&-8.84&852.96 $\pm$ 0.04 \\ \hline
MFH&41/2&-8.84&852.97 $\pm$ 0.04 \\ \hline
\end{tabular}
\caption{\label{tablesm1}
    DMC total energies ordered by lowest to highest with respect to $V_{0}$ values.
}
\end{table}

\subsection{QMC large flake size results}
\begin{figure}[h]
\centering
\includegraphics[width=0.750\columnwidth]{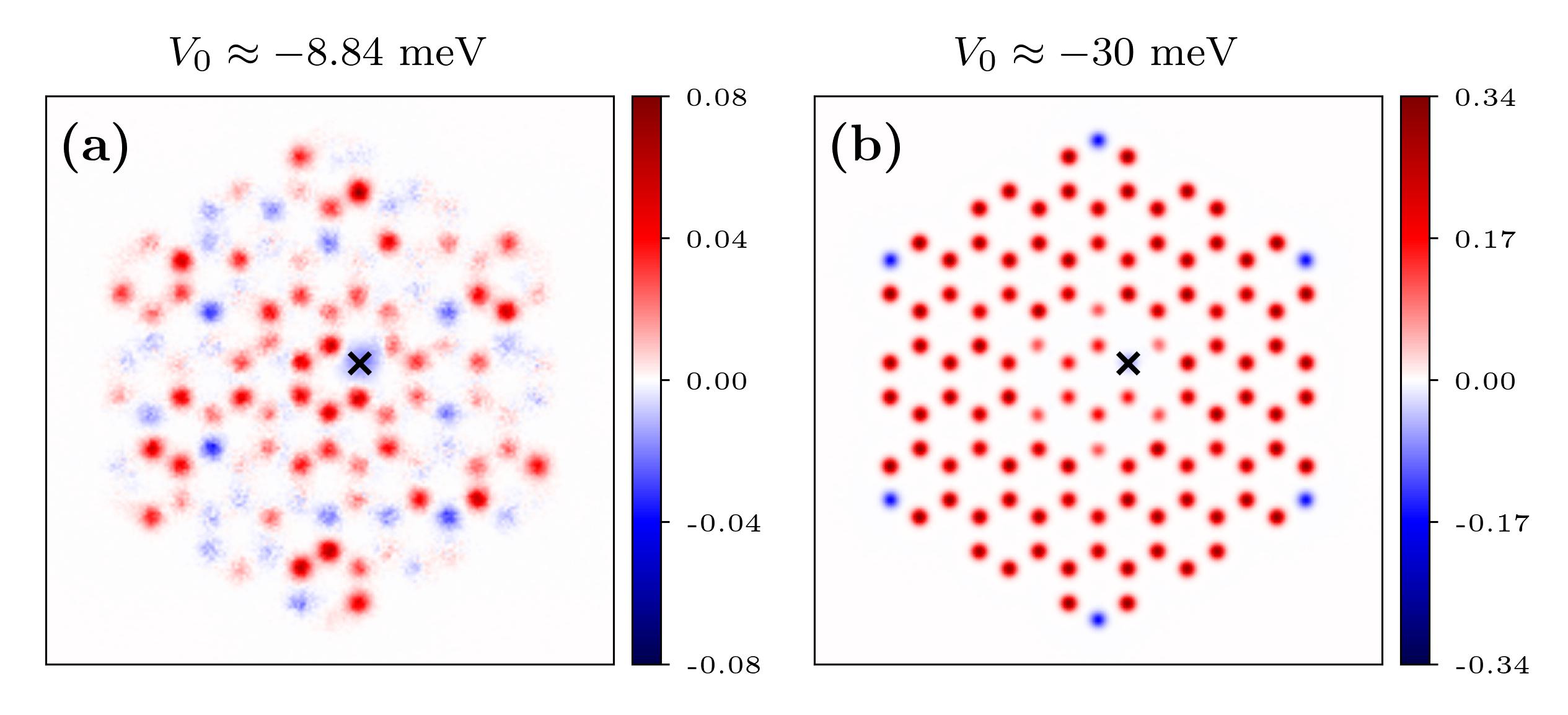}
\caption{\label{figuresm4}
    QMC pair spin density results for $N_{e}=113$ electrons in $N=114$ sites 
    from full $S_{z}$ scan. The results are ensemble averages at $T=4$ K since
    the total energies obtained from all $S_{z}$ values are similar within error bars.
    (a) $V_{0} \approx -8.84$ meV, $g \approx 0.0088$, $\langle S_{z} \rangle \approx 5.95$.
    (b) $V_{0} \approx -30$ meV, $g \approx 0.79$, $\langle S_{z} \rangle \approx 50.37$.
}
\end{figure}

In order to analyze another flake size, additional QMC calculations are performed for
$N_{e}=113$ electrons in $N=114$ sites for a hexagonal armchair flake. All $S_{z}$ values
are scanned for two potential depth $V_{0}$ values since full many-body calculations of
$113$ electrons are computationally expensive. In Fig.\ \ref{figuresm4}, QMC pair spin
density results are shown for (a) $V_{0} \approx -8.84$ meV, and (b) $V_{0} \approx -30$ meV.
$V_{0}$ values are chosen using $S_{z}$ scan results of $N=41$ electrons (a value around the start of transition in a liquid-like state, and around the end of the transition in a localized state). For both potential depth values, the $S_{z}$ scan of the total
energies result in similar values within error bars. Since the ground state cannot be
determined, an ensemble average is calculated using Boltzmann distribution at $T = 4$ K.
In Fig.\ \ref{figuresm4}(a),
$g \approx 0.0088$ and $\langle S_{z} \rangle \approx 5.95$; $g$ points out a metallic state, and $\langle S_{z} \rangle$ has a low value, as predicted. For Fig.\ \ref{figuresm4}(b); $g$ is closer to the fully polarized phase, as $g \approx 0.79$ and $\langle S_{z} \rangle \approx 50.37$. These results are consistent with those obtained for the $42$-site system, suggesting a possible transition from a metallic to a ferromagnetic phase with respect to $V_{0}$ in the $114$-site case.

